\documentclass[a4paper,11pt]{article}
\usepackage{pos}

\title{Learning Trivializing Flows in a $\phi^{4}$ theory from coarser lattices}

\author*[a]{David Albandea}
\author[b]{Luigi Del Debbio}
\author[a]{Pilar Hernández}
\author[b]{Richard Kenway}
\author[b]{Joe Marsh Rossney}
\author[a]{Alberto Ramos}

\affiliation[a]{Instituto de Física Corpuscular (CSIC -- University of
Valencia), Parque Científico, C/Catedrático José Beltrán, 2, 46980, Paterna,
Valencia, Spain}
\affiliation[b]{Higgs Centre for Theoretical Physics, School of Physics and
Astronomy, The University of Edinburgh, Edinburgh EH9 3FD, UK}

\emailAdd{david.albandea@ific.uv.es}

\abstract{The so-called trivializing flows were proposed to speed up Hybrid
    Monte Carlo simulations, where the Wilson flow was used as an approximation
    of a trivializing map, a transformation of the gauge fields which
    trivializes the theory. It was shown that the scaling of the computational
    costs towards the continuum did not change with respect to HMC. The
    introduction of machine learning tecniques, especially normalizing flows,
    for the sampling of lattice gauge theories has shed some hope on solving
    topology freezing in lattice QCD simulations. In this talk I will present
    our work in a $\phi^{4}$ theory using normalizing flows as trivializing
    flows (given its similarity with the idea of a trivializing map), training
    from a trivial distribution as well as from coarser lattices, and study its
    scaling towards the continuum, comparing it with standard HMC.}

\FullConference{%
  The 40th International Symposium on Lattice Field Theory (Lattice2023),\\
  July 31st - August 4th, 2023 \\
  Fermi National Accelerator Laboratory
}

\begin{document}
\maketitle

\section{Introduction}\label{sec:introduction}

\subsection{Normalizing flows and Critical Slowing Down}

Normalizing flows are a machine learning sampling technique introduced to lattice
field theories in \cite{Albergo2019} for a $\phi^{4}$ theory. There, they use a
neural network $f$ that generates field configurations $\phi$ following a model
distribution $p_{f}(\phi)=r(f(\phi))\left| \det \partial f(\phi) / \partial \phi
\right|$, by taking configurations $z=f(\phi)$ from a trivial probability
distribution $r(z)$ as input. The network is trained so that $p_{f}$ resembles
the probability distribution of the theory of interest, $p(\phi) =
e^{-S(\phi)}/Z$, where $S(\phi)$ is the action of the theory. The training of
the network usually consists on the minimization of the reverse Kullbach-Leibler
(KL) divergence between the network model distribution $p_{f}$ and the target
distribution $p$,
\begin{align}
D_{\text{KL}}(p_{f} \mid\mid p) = \int \mathcal{D}\phi \;
p_{f}(\phi) \log \frac{p_{f}(\phi)}{p(\phi)}.
\end{align}
This object satisfies $D_{\text{KL}}(p_{f} \mid \mid p) \ge 0$ and
$D_{\text{KL}}(p_{f} \mid \mid p) \Leftrightarrow p_{f} = p$, thus defining a
statistical distance between the two distributions. After its minimization, the
network is used as a proposal distribution in a Metropolis--Hastings algorithm
to obtain a Markov chain of configurations following $p(\phi)$.

One of the main results of \cite{Albergo2019} is that autocorrelation times of
the generated Markov chain do not scale when taking the continuum limit if the
neural networks are trained up to the same reference acceptance. This would
avoid the critical slowing down problem of local update algorithms, such as
Hybrid Monte Carlo (HMC), where autocorrelations grow towards the continuum with
the correlation length of the system as $\xi^2$. However, if one wants to study
the scaling of the total computational cost of the algorithm, one needs to
analyze the training costs as well and, as it was shown in \cite{DelDebbio2021}
for the same toy theory, the cost of keeping a reference Metropolis acceptance
of 70\% seems to scale approximately as $\sim \xi^{8}$, indicating a transfer of
the critical slowing down problem from the production of configurations to the
training cost of the networks.

\subsection{Flow HMC training from a trivial probability distribution}

In \cite{Albandea:2023wgd} we studied if one could benefit from normalizing
flows keeping the training costs as low as possible by using minimal network
architectures with few trainable parameters. Our idea was to use Lüscher's
trivializing flows algorithm in \cite{LuscherTrivializingMaps}, so that we can
use the normalizing flows to \emph{help} the HMC algorithm, rather than
replacing it.  For example, let us consider the partition function of our target
theory,
\begin{align}
    Z = \int \mathcal{D}\phi\; e^{-S(\phi)}.
\end{align}
One can use our trained network $f$ to make a change of variables $\tilde{\phi}
= f(\phi)$ so that the partition function becomes
\begin{align}
Z = \int_{ }^{ } \mathcal{D} \tilde{\phi }\; e^{-S[f^{-1}(\tilde{\phi })] + \log \det J[f^{-1}(
\tilde{\phi } )]} \equiv \int_{ }^{ } \mathcal{D}\tilde{\phi} \;
e^{-\tilde{S}(\tilde{\phi})},
\end{align}
where we have defined the new action $\tilde{S}( \tilde{\phi} ) \equiv
S[f^{-1}(\tilde{\phi })] - \log \det J[f^{-1}( \tilde{\phi } )]$. If the
Jacobian cancels out part of the action, then the probability distribution
$e^{-\tilde{S}( \tilde{\phi} )}$ might be easier to sample from than $e^{-S(
\phi )}$, and using HMC with the new action $\tilde{S}$ might yield lower
autocorrelation times. The workflow of the algorithm, which we call flow HMC
(FHMC), would then be
\begin{enumerate}
    \item
        Train the network $f$ by minimizing the KL divergence.
    \item
        Run the HMC algorithm to build a Markov chain of configurations
        following $\tilde{p}(\tilde{\phi}) = e^{-\tilde{S}(\tilde{\phi})}$,
        \begin{align*}
            \{ \tilde{\phi}_{1},\; \tilde{\phi}_{2},\; \tilde{\phi}_{3},\; \dots ,\;
            \tilde{\phi}_{N} \} \sim e^{-\tilde{S}(\tilde{\phi})}.
        \end{align*}
    \item
        Apply the inverse transformation $f^{-1}$ to every configuration in the
        Markov chain to undo the variable transformation.  This way we obtain a
        Markov chain of configurations following the target probability
        distribution $p(\phi) = e^{-S[\phi]}$,
        \begin{align*}
            \{ f^{-1}(\tilde{\phi}_{1}),\; f^{-1}(\tilde{\phi}_{2}),\;
            f^{-1}(\tilde{\phi}_{3}),\; \dots ,\; f^{-1}(\tilde{\phi}_{N}) \} =
            \{ \phi_{1},\; \phi_{2},\; \phi_{3},\; \dots ,\; \phi_{N} \} \sim
            e^{-S(\phi)}.
        \end{align*}
\end{enumerate}
The important point is that the acceptance of this algorithm depends mainly on
how well one integrates the HMC equations of motion. This means that the
algorithm will work, no matter how well one trains the network $f$.

Lüscher proposed this algorithm using the Wilson flow as an approximate
trivializing map \cite{LuscherTrivializingMaps}, but it was not good enough to
improve the scaling of autocorrelation times towards the continuum in a
CP$^{N-1}$ theory with topology \cite{Schaefer2011}. The hope is that
normalizing flows can play a better role as approximate trivializing maps.

\begin{figure}[t]
	\centering
    \includegraphics[width=0.6\linewidth]{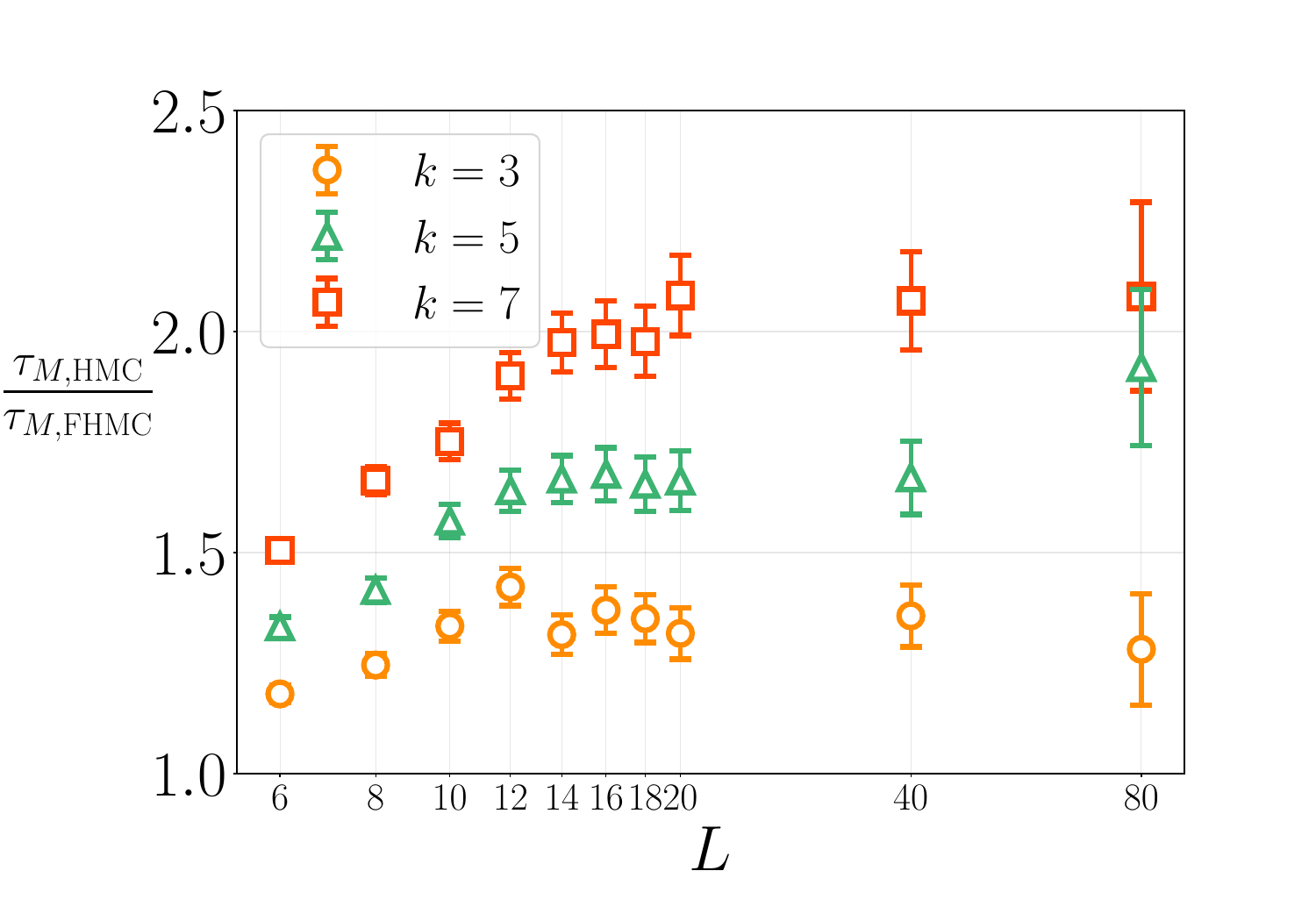}
    \caption{Continuum scaling, keeping $L / \xi = 4$, of the ratio of
        autocorrelation times of the magnetization of HMC over FHMC, for
        networks trained from a normal distribution and with kernel sizes
    $k=3,5,7$.}
        \label{fig:tauMratio-fixedarch}
\end{figure}

A very similar idea has already been tested in \cite{Jin2022}, where they
minimize the HMC force instead of the KL divergence. Also, in
\cite{Albandea:2023wgd} we focus on the scaling of autocorrelation times using
cheap training setups, with a network architecture with a single affine coupling
layer \cite{dinh2017density} and no hidden layers.  The application of the
network layers on a configuration is
\begin{align}
    \phi \to e^{-|s(\phi)|} \odot \phi + t(\phi),
\end{align}
with $s(\phi)$ and $t(\phi)$ being convolutional neural networks with kernel
size $k$. The main result of \cite{Albandea:2023wgd} is shown in
Fig.~\ref{fig:tauMratio-fixedarch}, where we plot the scaling of the ratio
autocorrelation times of the magnetization $M=\frac{1}{V}\sum_{i}^{}\phi_{i}$ of
HMC and FHMC for kernel sizes $k=3,5,7$, finding that the scalings of both
algorithms are the same if the network architecture is kept fixed. In the
following we explore two possible solutions for this.

\section{Training from a coarser theory}
\label{sec:org0af4a55}

A possible explanation of why keeping a fixed architecture implies the scaling
of FHMC is not improved with respect to HMC is that the footprint of the network
is constant in lattice units, and therefore decreases in physical units as we
approach to the continuum. A possible solution, which we already explored in
\cite{Albandea:2023wgd}, is to scale the footprint of the networks with the
correlation length of the system. Another possible solution, which is the focus
of this study, is to change the input theory distribution as we approach the
continuum limit.

The input distribution \(r(\phi)\) does not need to be one from which sampling is
trivial in order for the method to be useful for the sampling of the target
distribution at a given coupling \(\beta\),
\begin{align}
    p_{\beta}(\phi) = \frac{1}{Z_{\beta}} e^{-S_{\beta}(\phi)}.
\end{align}
It is natural to think that samples from the target theory itself at a different
coupling value, $p_{\beta'}(\phi)$, are a better approximation to the target
distribution. Although being potentially more costly, it is expected that using
this distribution for the training of the network would lead to a faster
minimization of the KL divergence and, generally, to a better mapping between
the two probability distributions.

The choice of the coupling for the input distribution should be chosen so that
it is significantly easier to sample, using traditional methods, than the target
distribution. That is, the input theory should be a coarser theory, and the
application of the network \(f\) on samples from this theory after training is
expected to reduce the lattice spacing, thus building a sort of inverse
renormalization group transformation\footnote{ See also R. Abbot's talk, \emph{
Multiscale Normalizing Flows for Gauge Theories}, and N. Matsumoto's talk,
\emph{Decimation map in 2D for accelrating HMC}, in LATTICE2023 for similar
approaches for lattice gauge theories.} \cite{Bachtis:2021eww}. As a proof of
concept, we will always choose the coupling of the input distribution such that
the lattice spacing is halved under the application of the network \(f\), i.e.
$a(\beta) = a'(\beta') / 2$.  Additionally, and following the same strategy as
in the previous work, we focus on a minimal model with only one affine coupling
layer to reduce training costs as much as possible (see \cite{Albandea:2023wgd}
for more details).

In principle, the good thing about this is that the longest correlation length,
which is the important one for the HMC evolution, is already captured in the
coarse theory, $p_{\beta'}$. The problem though is that one cannot train
directly at fixed physical volume, because the number degrees of freedom in the
coarse and fine theories would not be the same. We have tried two possible
workarounds for this.

\subsection{Training from a coarser theory with bigger physical volume}
\label{sec:org10c99d6}

\begin{figure}[t]
	\centering
    \includegraphics[width=0.6\linewidth]{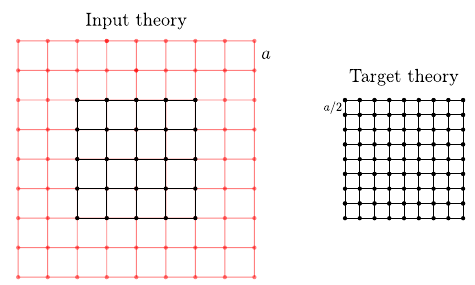}
    \caption{Sketch of configuration from the input theory and target theory
        when training from a coarser theory with bigger physical volume. The
        black lattice points represent a same physical volume in both input and
        target theories, while the red points have been added to the input
        theory so that the number of degrees of freedom of both theories match.}
    \label{fig:training-bigger-physical-volume}
\end{figure}

The simplest option to match the number of degrees of freedom in both input and
target theories is to increase the physical volume of the input theory. This is
depicted in Fig.~\ref{fig:training-bigger-physical-volume}, where the region in
black in both input and target theories denote an equivalent physical volume,
while the red points have been added to match the number of lattice points.

The KL divergence between the model and target distribution is
\begin{align}
D_{\text{KL}}(p_{f} \mid\mid p_{\beta(a)}) = \int \mathcal{D}\phi \;
p_{f}(\phi) \log \frac{p_{f}(\phi)}{p_{\beta(a)}(\phi)},
\end{align}
with the model distribution $p_{f}(\phi) = p_{\beta'(a')}(f(\phi)) \left| \det
\frac{\partial f(\phi)}{\partial \phi} \right|$.  Since the normalization
constants of none of the distributions is known, we minimized instead the loss
function
\begin{align}
L = Z_{\beta'} \left[ D_{\text{KL}} + \log \frac{Z_{\beta'}}{Z_{\beta}} \right],
\end{align}
whose unknown minimum is \(Z_{\beta'} \log
\frac{Z_{\beta'}}{Z_{\beta}}\). For this minimization one needs to
stochastically estimate the loss function by drawing samples \(\phi_{i} \sim
p_{f}(\phi)\) from the model, for which one needs to draw samples from the input
distribution \(p_{a'(\beta')}\) first using traditional methods. If using the
standard HMC algorithm, in order to have an unbiased estimator of the loss
function one would need to select the samples separated by 2 times the
largest autocorrelation time at \(\beta'\), which is expected to be 4 times
cheaper than it would be at the coupling constant of the target theory,
\(\beta\).

\begin{figure}[!t]
    \begin{minipage}{\textwidth}
        \centering
        \begin{minipage}{0.32\textwidth}
            \centering
            \begin{tabular}{lr}
                Algorithm & \(\tau_M\)\\
                \hline \\
                HMC & 77.9(1.5)\\ \\
                FHMC & 63.6(2.2)\\ (bigger volume) \\ \\
                FHMC  & 56.9(1.8)\\ (from Gaussian) \\ \\
                FHMC  & 32.5(1.2)\\ (4-config. interp.) 
            \end{tabular}
            \captionof{table}{Autocorrelation times for HMC and FHMC at $L=16$
                and $\beta=0.634$. The networks used for FHMC were trained from:
                a normal distribution; a coarser theory with a bigger
                physical volume; and a 4-configuration interpolation from a
            coarser theory with the same physical volume.}
            \label{tab:mainresults}
        \end{minipage}\hfil
        \begin{minipage}{0.59\textwidth}
            \centering
            \includegraphics[width=\linewidth]{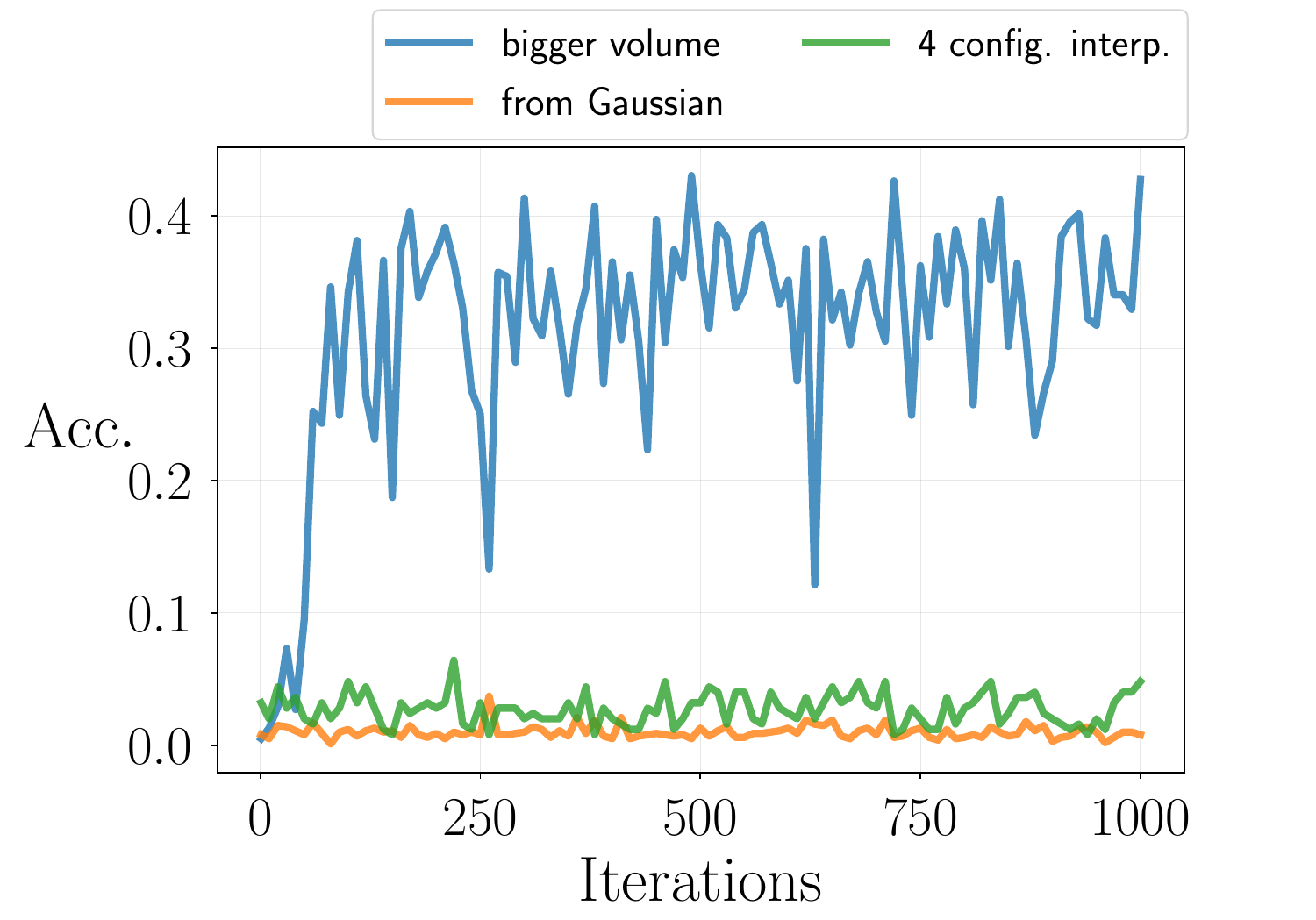}
            \caption{Evolution with the training iterations of the acceptance of
                networks trained from a normal distribution (orange), from a
                coarser theory with a bigger physical volume (blue) and from a
                4-configuration interpolation from a coarser theory with the
            same physical volume (green).}
            \label{fig:acc-evolution-fhmc}
        \end{minipage}\hfill
    \end{minipage}
\end{figure}

In Fig.~\ref{fig:acc-evolution-fhmc} we plot the evolution of the Metropolis
acceptance of different networks as a function of the number of training
iterations, with a target theory with $L=16$ and $\beta=0.634$, parameters which
were chosen so that $L / \xi = 4$. The case in which the training is performed
directly from independent and normal random numbers as input distribution is
shown in orange and yields a low Metropolis acceptance (of the order of 1\%) as
was already studied in \cite{Albandea:2023wgd}. On the other hand, the blue
curve corresponds to a network trained from a theory with $L=16$ and
$\beta=0.576$, which has twice the lattice spacing of the target theory but also
a bigger physical volume, $L / \xi = 8$ (and therefore is not in the same line of
constant physics), as is needed to match the number of degrees of freedom of
both theories. One can see that the acceptance saturates after a few hundreds of
iterations due to the simplicity of the architecture, and that it reaches a much
higher acceptance than using independent normal random numbers as input
distribution.

However, when using the networks as a transformation of variables for the FHMC
algorithm, the opposite happens: in Tab.~\ref{tab:mainresults} we see that
although the network trained from the theory with double the lattice spacing and
bigger physical volume yields a magnetization autocorrelation time of
$\tau_{M}=63.6(2.2)$, thus improving the autocorrelation time of standard HMC,
$\tau_{M}=77.9(1.5)$, it is not better than using a network trained directly
from a normal input distribution, which leads to $\tau_{M}=56.9(1.8)$.

This may be an indication that the Metropolis acceptance of the network is not
the best metric to assess if the trained network is a good transformation for
the FHMC algorithm, and that the minimization of the KL divergence is probably
not the best loss function for the optimization of the network.

\subsection{Training from an interpolation of 4 coarser configurations at a same
physical volume}
\label{sec:org56003d3}

\begin{figure}[t]
	\centering
    \includegraphics[width=0.6\linewidth]{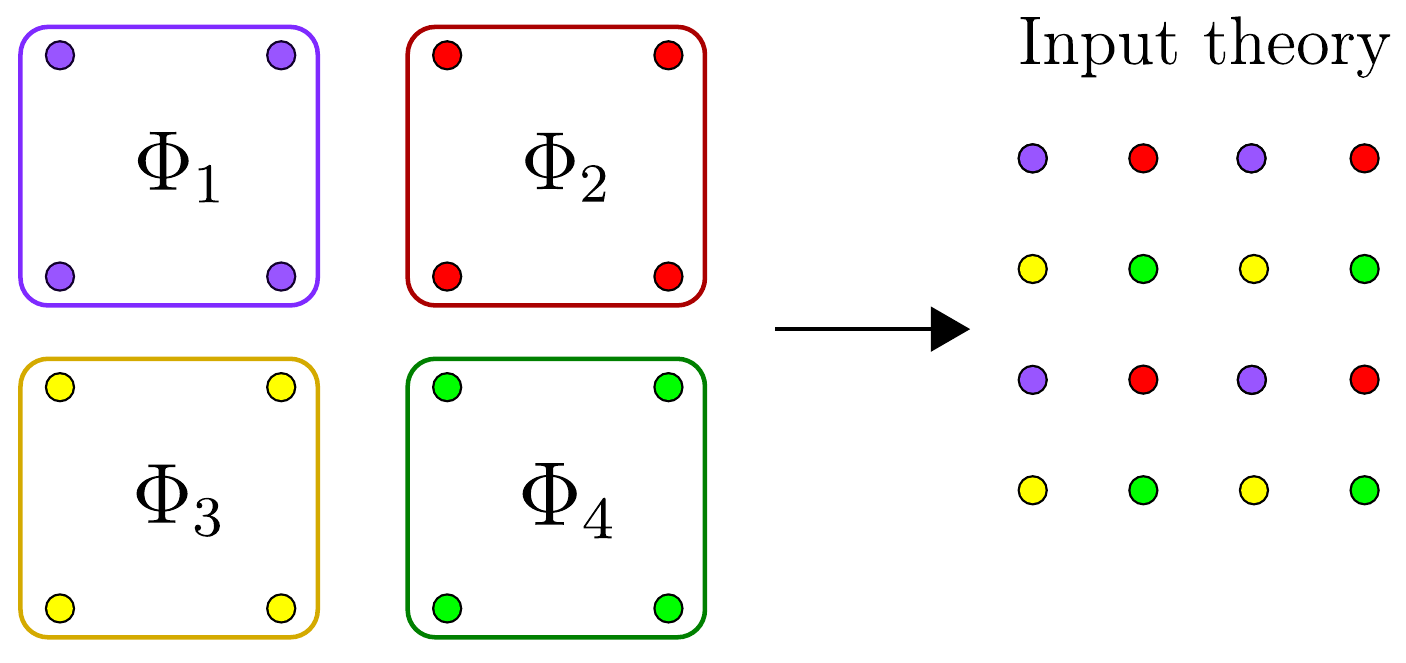}
    \caption{Sketch of the combination of four coarse configurations
    $\{\Phi_{i}\}_{i=1}^{4}$ into a configuration $\tilde{\phi}$ as showed in
Eq.~\ref{eq:4config-combination}.}
    \label{fig:4confinterp}
\end{figure}

A reason why training from a bigger physical volume might not be good enough is
that the correlation length in lattice units of the two systems is different,
and this needs to be learnt by the network. An alternative approach to match the
number of degrees of freedom of both theories, more loyal to the idea of the
inverse renormalization group, is to train at fixed physical volume, so that the
presence of the correlation length in the input theory is reinforced. For this
we combined 4 different configurations \(\{\Phi_{i}\}_{i=1}^{4}\) from the
coarse theory into a bigger configuration \(\tilde{\phi}\), thus defining the input
distribution as
\begin{align}
\tilde{p}(\tilde{\phi}) =
p_{\beta'(a')}(\Phi_{1})\; 
p_{\beta'(a')}(\Phi_{2})\;
p_{\beta'(a')}(\Phi_{3})\;
p_{\beta'(a')}(\Phi_{4})
= \frac{1}{\tilde{Z}} e^{-\sum_{i}^{}S_{\beta'(a')}(\Phi_{i})},
\end{align}
with each constituent configuration \(\Phi_{i}\) being sampled from the coarse
theory, $p_{\beta'(a')}(\Phi_{i}) = \frac{1}{Z_{\beta'}}
e^{-S_{\beta'(a')}(\Phi_{i})}$.
As also shown in Fig.~\ref{fig:4confinterp}, the four different configurations
are combined into a new configuration \(\tilde{\phi}\) such that
\begin{align}
\tilde{\phi}(2n_{x}, 2n_{y}) &= \Phi_{1}(n_{x},n_{y}),\nonumber\\
\tilde{\phi}(2n_{x} + 1, 2n_{y}) &= \Phi_{2}(n_{x},n_{y}),\nonumber\\
\tilde{\phi}(2n_{x}, 2n_{y} + 1) &= \Phi_{3}(n_{x},n_{y}),\nonumber\\
\tilde{\phi}(2n_{x} + 1, 2n_{y} + 1) &= \Phi_{4}(n_{x},n_{y}).
\label{eq:4config-combination}
\end{align}
with $n_{x}, n_{y}=0, \dots, L-1$. Under the application of a network $f$ on the
newly built configuration, $\phi = f^{-1}(\tilde{\phi})$, the model distribution
becomes $p_{f}(\phi) = \tilde{p}(f(\phi)) \left| \det \frac{\partial
f(\phi)}{\partial \phi} \right|$.

The evolution of the Metropolis acceptance during the training of this network
is displayed in the green curve of Fig.~\ref{fig:acc-evolution-fhmc}, where the
input configurations $\Phi_{i}$ are sampled from a theory with $L=8$ and
$L=0.576$ with twice the lattice spacing of the target theory, as was the case
in the previous section; however, now these configurations lie in the same line
of constant physics as the target theory, $L / \xi = 4$. One can see that this
leads to an acceptance comparable to training directly from independent normal
distributions, much lower than the one achieved by the network trained from a
bigger physical volume as studied in the previous section. However, as is shown
in Tab.~\ref{tab:mainresults}, when used as a transformation of variables for
FHMC it leads to the lowest autocorrelation time, indicating again that a higher
Metropolis acceptance of the network does not imply lower autocorrelation times
for the FHMC algorithm, and that reinforcing the same correlation length and
physical volume in both input and target theories plays a more important role.

\begin{figure}[t]
	\centering
    \includegraphics[width=0.6\linewidth]{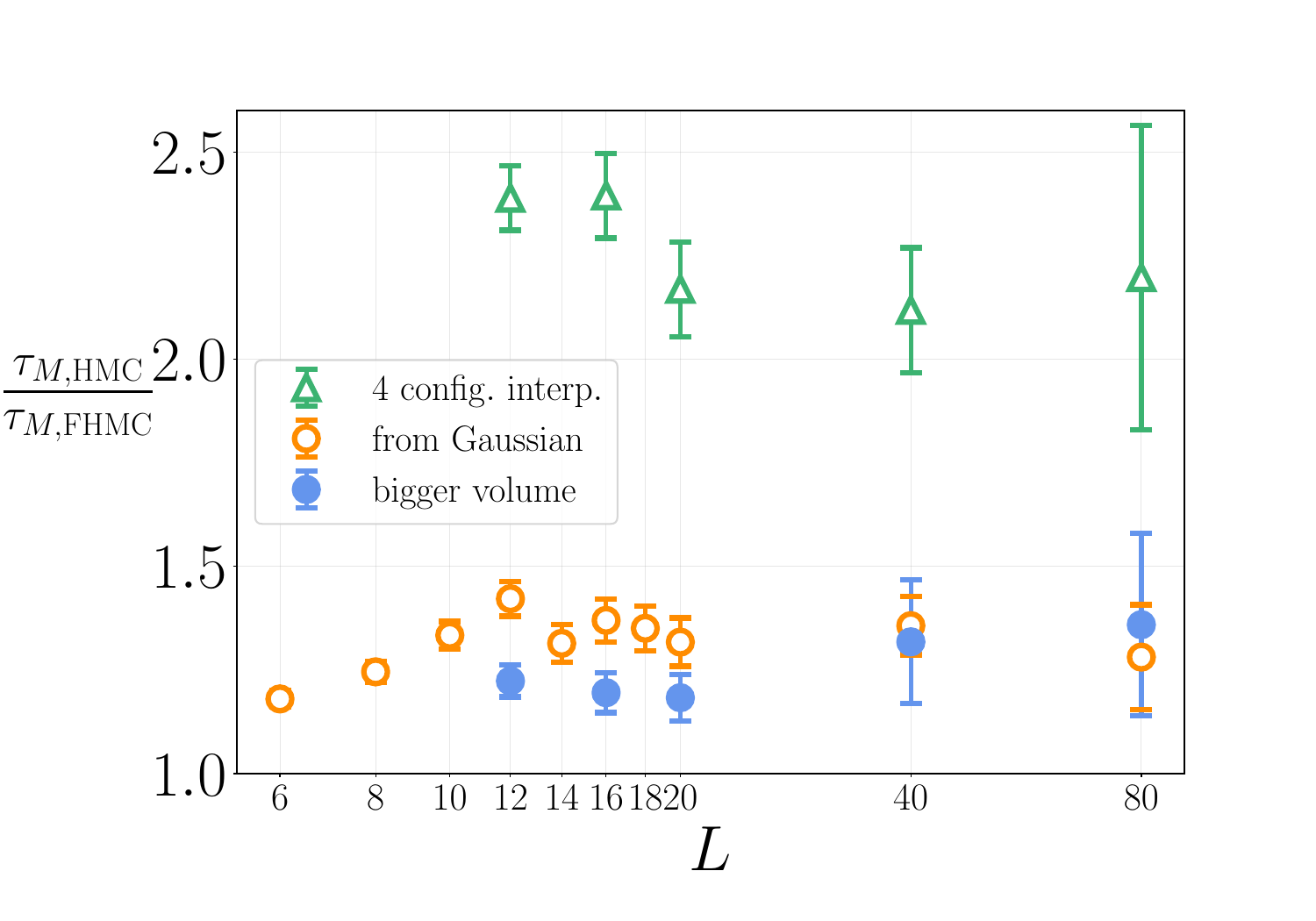}
    \caption{Continuum scaling, keeping $L / \xi = 4$, of the ratio of
        autocorrelation times of the magnetization of HMC over FHMC, for
        networks trained from a normal distribution (open orange circles), from
        a coarser theory with a bigger physical volume (full blue circles) and
        from a 4-configuration interpolation from a coarser theory with the same
        physical volume (green triangles).}
    \label{fig:coarse-training-scaling}
\end{figure}

\section{Scaling}
\label{sec:orgcfba2e1}

Fig.~\ref{fig:coarse-training-scaling} shows again the continuum scaling of the
ratio of autocorrelation times of the magnetization of HMC over FHMC, this time
for networks trained from a normal distribution (open orange circles), from a
coarser theory with a bigger physical volume (full blue circles) and from a
4-configuration interpolation from a coarser theory with the same physical
volume (green triangles).  Although autocorrelation times are improved with
respect to HMC, the scaling seems to be the same towards the continuum for a
fixed network architecture.

\section{Conclusions}

We have further studied the FHMC algorithm, which we introduced in
\cite{Albandea:2023wgd} as a proposal to improve the continuum scaling of HMC,
in a $\phi^{4}$ toy theory.  The algorithm uses normalizing flows as approximate
trivializing maps for the Lüscher algorithm proposed in
\cite{LuscherTrivializingMaps}, focusing on cheap training setups to avoid the
bad scaling of the training costs of normalizing flows. 

Knowing that training from a normal distribution and keeping a fixed network
architecture towards the continuum does not lead to a better scaling with
respect to HMC, we have trained instead from the theory of interest but at a
coarser value of the coupling. We have found that training from a coarser theory
at a bigger physical volume leads to networks with much higher
Metropolis--Hastings acceptances, but worse autocorrelation times when used as a
transformation of variables for the FHMC algorithm, indicating that the
minimization of the KL divergence is probably not the best optimization
method for this algorithm.

We have also found that a 4-configuration interpolation of the coarser input
theory with the same physical volume as the target theory leads to the lowest
autocorrelation time, indicating that reinforcing the presence of the target
correlation length in the input theory and having the same physical volume in
both input and target theories plays an important role in the algorithm.
Although this seems to have the same scaling towards the continuum as HMC, a
possible application in which the input theory is iterated to coarser and
coarser lattice spacings following the procedure above could nonetheless
improve the scaling towards the continuum, and its study is left for future
work.

\section*{Acknowledgments}
\addcontentsline{toc}{section}{Acknowledgements}

We acknowledge support from the Generalitat Valenciana grant PROMETEO/2019/083,
the European projects H2020-MSCA-ITN-2019//860881-HIDDeN and
101086085-ASYMMETRY, and the national project PID2020-113644GB-I00. AR
acknowledges financial support from Generalitat Valenciana through the plan GenT
program (CIDEGENT/2019/040). DA acknowledges support from the Generalitat
Valenciana grant ACIF/2020/011. JMR is supported by STFC grant ST/T506060/1. LDD
is supported by the UK Science and Technology Facility Council (STFC) grant
ST/P000630/1.

We also acknowledge the computational resources provided by Finis Terrae II
(CESGA), Lluis Vives (UV), Tirant III (UV). The authors also gratefully
acknowledge the computer resources at Artemisa, funded by the European Union
ERDF and Comunitat Valenciana, as well as the technical support provided by the
Instituto de Física Corpuscular, IFIC (CSIC-UV).



\bibliography{references}
\addcontentsline{toc}{section}{References}
\bibliographystyle{unsrt}

\end{document}